# Role of Zr in Cu-rich single-phase and nanocomposite Cu-Zr: molecular dynamics and experimental study

J. Houska*, M. Zhadko, R. Cerstvy, D. Thakur, P. Zeman

*Department of Physics and NTIS - European Centre of Excellence, University of West Bohemia, Univerzitni 8, 301 00 Plzen, Czech Republic*

*corresponding author, email jhouska@kfy.zcu.cz

## Abstract

The non-equilibrium atom-by-atom growth of Cu-rich Cu-Zr thin films has been investigated by a combination of magnetron sputter deposition and molecular dynamics simulations. We focus on the role of Zr in the transition from large solid solution crystals through a nanocomposite (around ≈5 at.% Zr) to a metallic glass. We find, contrary to the assumption based on equilibrium phase diagram, that in this non-equilibrium case most of the grain refinement and most of the hardness enhancement (from 2.5-3 to 4-5 GPa) takes place in the compositional range (up to ≈3 at.% Zr) where many or even most Zr atoms (depending on the sputtering regime) are in the supersaturated solid solution rather than at the grain boundaries. The results are important for the design and understanding of technologically important nanostructured metallic films. In parallel, from the methodology point of view, the results include an early example of modelling the atom-by-atom nanocomposite growth.

## 1. Introduction

There are worldwide efforts to master the preparation of thin films of all kinds of technologically important materials: from single-phase crystalline through nanocomposite to amorphous, from covalent through metal oxides or nitrides to metals, and many even more complex ones. This includes the research in the field of thin films of Cu-Zr, both Cu-rich crystalline [1-3] and metallic glasses [4-6], in the first place in the context of their mechanical properties. The research builds on the results obtained for bulk Cu-Zr [7,8] and for thin films of other binary Cu-based alloys [9,10]. Magnetron sputtering [11], either conventional or high-power impulse (HiPIMS), is a universal, scalable and probably the most frequent technique used to deposit thin films including the atom-by-atom growth of Cu-Zr. Owing to its non-equilibrium nature, it allows one to prepare both thermodynamically stable and metastable structures.

The experimental research in the field of thin films is increasingly frequently supported by molecular dynamics (MD) simulations of their atom-by-atom growth [12]. The strengths of this approach are spanning from time and cost savings through disentanglement of phenomena which take place in parallel in the experiment (conditions for crystal nucleation and for crystal growth [13], effects of impacts of atoms of various elements with various energies [14], energy delivered by ion bombardment and by ohmic heating [15]) to providing exact atomic structures and medium-range order of amorphous networks [16]. The similarity of the simulation protocol which basically reproduces what is happening experimentally makes the results relatively easy to interpret and directly useful. Examples of previous successful usage of these simulations are ranging from covalent C [17], Si [18], SiC [19] or SiNH [20] through ionic $TiO_2$ [13], $ZrO_2$ [14], MgO [21] or ZnO [22] to metallic Cu-Zr [23,24] or Cu-Zr-Al [16,25].

However, there is a knowledge gap resulting from the following. First, in the case of Cu-Zr (arguably a prototype of metallic glasses, MGs), the experimental and especially theoretical research has been primarily focused on characteristics of MGs in the central part of the composition range. The efforts devoted to the not yet established mechanism of grain refinement and hardening at only few at.% of Zr in thin films prepared by non-equilibrium techniques have been significantly lower. This may include grain boundary segregation (phenomenon of high overall interest [26,27], and here consistent with the almost zero solubility of Zr in fcc Cu according to the equilibrium phase diagram), formation of a Zr-rich amorphous phase [1,2]) and a supersaturated solid solution [3]. Second, because the simulation algorithm reproducing the atom-by-atom growth is not only exceptionally revealing but also (for classical MD) relatively slow, previous growth simulations of reasonably thick films of various materials were rather focused on the amorphous phase or single crystals in relatively narrow simulation cells than on nanocomposites. Due to the decreasing computational costs, the time is now ripe to enlarge the simulations cells and to capture also the growth of nanocomposites and heterogeneous structures in general by MD.

We contribute to filling this knowledge gap by the following. First, we prepared Cu-Zr films at Zr content, [Zr] (given in at.% throughout this paper), below 5% by two versions of magnetron sputtering. While a detailed study of film characteristics will be published separately, here we report those characteristics which are crucial in the present context: stress-free lattice parameter of Cu crystals (measure of the presence of Zr inside them) and hardness. Second, we analyze the role of Zr in the transition from solid solution through a nanocomposite to a metallic glass using large-scale growth simulations. Both sets of results are complementary to our previous experiments (lowest non-zero [Zr] of 12%) and smaller-scale growth simulations focused of glass-forming Cu-Zr compositions [24]. As the results include an early example of modelling the nanocomposite growth (more revealing than simulations which cannot directly reproduce the present experiments such as liquid-quench [28-31] or Monte-Carlo [32] or Voronoi construction [33]), considerations related to its methodology are included as well.

## 2. Methodology

*1. Experiment*

Two film series were deposited using different versions of magnetron sputtering. In the first series, nanocrystalline Cu and Cu-Zr films were prepared by direct current (DC) magnetron co-sputtering of separate Cu and Zr targets with [Zr] ranging from ≈0.4 to 2.7%. In the second series, the magnetron with the Cu target was operated in the HiPIMS regime while the magnetron with the Zr target was operated in the DC regime, and [Zr] was in the range from 1.0 to 4.4%. This second series was prepared under high-density discharge conditions [4]. All films were deposited under fixed conditions, except varying [Zr] by adjusting the power on the Zr target. The films were deposited onto ultrasonically pre-cleaned glass substrates in pure Ar at a pressure of 0.533 Pa (4 mTorr) without substrate bias and external heating. The base pressure before each deposition was lower than $5 \cdot 10^{-5}$ Pa. The target-to-substrate distance was set to 150 mm.

The elemental composition of the as-deposited films was measured by wavelength dispersive spectroscopy (WDS) in a scanning electron microscope (Su-70, Hitachi) with the primary electron energy of 15 kV. For the quantitative analysis, Cu and Zr standards were used. Before the WDS analysis a thin layer of carbon with the thickness of 14 nm was sputtered on the film surface. The error of the elemental analysis was determined to be 0.1 at.%.

The structure of as-deposited films was characterized by X-ray diffraction (XRD) using a PANalytical X'Pert PRO MPD diffractometer with a CuK$\alpha$ ($\lambda$ = 0.154187 nm) radiation. The measurements were carried out both in the Bragg-Brentano geometry (used to evaluate the grain size) and at a glancing incidence of 4° (GIXRD; used to evaluate the strain-free lattice parameter). All samples were scanned over the 2$\theta$-range from 20° to 120° with a scanning speed of 0.1 °/s (Bragg-Brentano) or 0.017 °/s (GIXRD). The data were processed by a PANalytical software package HighScore Plus. The grain size (size of coherently diffracting Cu regions) was estimated using the Scherrer equation from the full width at half maximum of the Cu (111) diffraction peak corrected for instrumental broadening by an NIST LaB$_6$ powder standard.

The hardness of the films was measured at room temperature using a Fischerscope H100 microhardness tester, equipped with a Vickers diamond indenter. Minimum 20 indents were made on each film with a consistent load of 10 mN, and the results were averaged. In all cases, the indentation depth was significantly less than 10% of the film thickness.

*2. Simulations*

Molecular-dynamics growth simulations of Cu-rich Cu-Zr ([Zr] ≤ 20%) were performed by the LAMMPS code [34]. The horizontal size of the simulation cell was approximately 22×22 nm, with periodical boundaries in horizontal directions and a vacuum slab in the vertical direction. The growth template was 16 nm thick crystalline Cu (001) (72 000 atoms, including 14 400 frozen atoms at the bottom). If we let alone the cell dimension (6×enhanced) and in turn square area of the growing surface (36× enhanced), the rest of the methodology follows our previous smaller-scale simulations dealing with Cu-Zr MGs [24]. The interatomic interactions were represented by the well established Mendelev force field [35] of the Finnis-Sinclair type (one of several Cu-Zr force fields developed by this team, differing mainly in treating glass-forming compositions which is beyond the present scope). Note also the recent comparison of Cu-Zr force fields developed by various teams [36], very successful from the Mendelev point of view. The temperature of the non-frozen substrate atoms was 300 K (NVT canonical ensemble), while the temperature of the film atoms was not controlled (NVE microcanonical ensemble). The energy of arriving atoms was 1 eV. At each [Zr], ≈200 000 atoms were deposited using a recursive simulation protocol. In each deposition step, 144 atoms arrived, sufficiently far from each other to start as isolated. The length of each deposition step was 1.5 ps (3000 MD steps with a length of 0.5 fs), ≈5× longer than the time of thermal spike cooling [24]. In the analysis of short- and medium-range order, the interatomic bonds (nearest neighbors) were identified using bonding distance cutoffs of 3.2 Å for Cu-Cu, 3.6 Å for Cu-Zr and 3.8 Å for Zr-Zr (see the pair correlation functions in [24]).

The analysis of long-range order is, in line with the present aim, focused on the unknown distribution of atoms of both elements in the binary materials, not on the known and visible crystal structure and orientation. Thus, it is not to be confused with algorithms based on, e.g, common neighbor analysis and used, e.g., to identify crystals in unary materials [37] or to characterize metallic glasses [16]. The Cu-based grains in the present sense of the word were defined as zones (not only crystals at low [Zr] but also amorphous zones at high [Zr]) fully surrounded by boundary atoms. Various definitions of boundary atoms were tested for this purpose, ranging from only Zr atoms to Zr atoms and their $1^{st}$-$6^{th}$ nearest neighbors. According to a comparison of automatically calculated data based on these definitions with a visual inspection, defining the boundary atoms as Zr atoms and their $1^{st}$ and $2^{nd}$ nearest neighbors is, by far, most meaningful. While this definition may sound relatively strict (as shown below, most atoms in a nanocomposite are actually boundary atoms), less strict definitions are prone to lead to a "hole" in a crystal surface and to treating several visually almost separated crystals as a single one.

## 3. Results and discussion

The measured stress-free lattice parameter of fcc Cu-based nanocrystals, $a_{fcc}$, is shown in Fig. 1a. Let us start by recalling the very different interatomic distances in close-packed fcc Cu and hcp Zr of 2.556 Å and 3.232 Å, respectively, corresponding to lattice parameters $a_{fcc\_Cu} = 2.556 \times 2^{1/2} = 3.615$ Å and $a_{fcc\_Zr} = 3.232 \times 2^{1/2} = 4.571$ Å. Thus, $a_{fcc}$ constitutes a sensitive measure of Zr content in the nanocrystals, $[Zr_{cr}] = (a_{fcc}^3 - a_{fcc\_Cu}^3)/(a_{fcc\_Zr}^3 - a_{fcc\_Cu}^3)$ if we conservatively average the atomic volumes (averaging the atomic sizes would lead to even higher $[Zr_{cr}]$ values).

In the case of DC sputtering of both Cu and Zr, leading to relatively low energy of arriving atoms and in turn growth conditions very far from the equilibrium, $a_{fcc}$ almost monotonically increases with [Zr]. At [Zr] ≤ 1% the increase is basically linear and almost equal to the corresponding weighted averages of atomic volumes. In other words, according to $a_{fcc}$ almost all Zr atoms end up in supersaturated Cu-Zr solid solution ($[Zr_{cr}] \approx [Zr]$) rather than at the grain boundaries. The increasing [Zr] in this range leads also to a steep grain refinement (inset in Fig. 1a), which has to be explained (at a negligible grain boundary segregation) by lower growth rate of Zr-containing crystals after their nucleation. At higher [Zr] the increase of $a_{fcc}$ slows down but continues, up to $a_{fcc} = 3.631$ Å indicating $[Zr_{cr}] = 1.3\%$ at the overall [Zr] = 2.3%.

In the case of HiPIMS of Cu combined with DC sputtering of Zr, the degree of ionization of the film-forming Cu flux is much higher and the energetic bombardment of growing films is much stronger, arguably leading to higher mobility of film atoms and to growth conditions relatively closer to the equilibrium. Consequently, the enhancement of $a_{fcc}$ is on the average lower ($a_{fcc}$ values obtained using the former technique constitute an upper envelope of $a_{fcc}$ values obtained using this technique), indicating solid solutions which are still qualitatively supersaturated but quantitatively contain on the average less Zr.

Our data smoothly connect to those reported (also using magnetron sputtering) by Chakraborty et al. [3], except that the trends yielded by our data mostly saturate between [Zr] ≈ 1% and 3% while the latter data exhibit an additional step up to $a_{fcc} = 3.641$ Å indicating $[Zr_{cr}] = 2.1\%$ Zr at the overall [Zr] = 4.5%. In any case note the consistency of the aforementioned maximum $[Zr_{cr}]$ values with those between 1.33% and 2.05% reported using a different technique (vacuum melting) by Tenwick et al. [38].

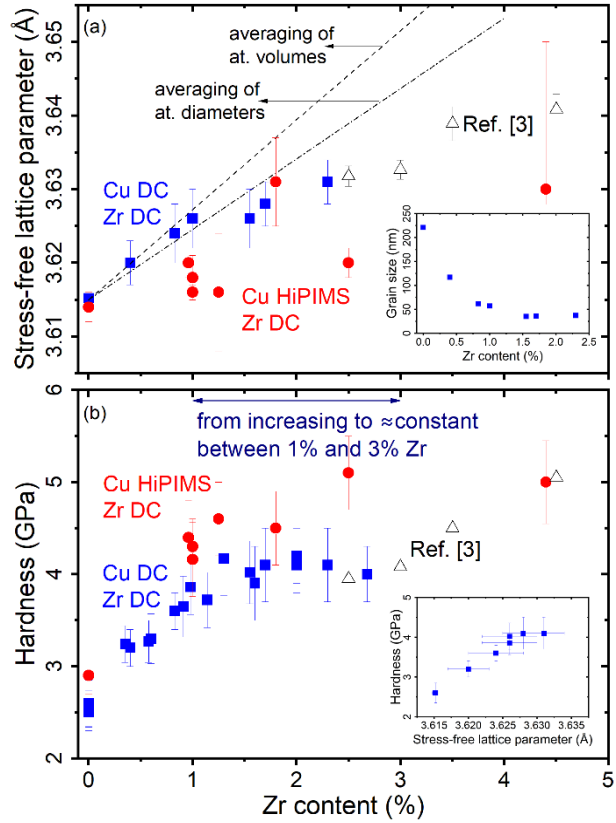

**Figure 1**: Measured stress-free lattice parameter (panel a) and hardness (panel b) of Cu-Zr in the compositional range where fcc crystals have been observed by XRD. There are results obtained by DC sputtering of both Cu and Zr (squares), results obtained by HiPIMS of Cu and DC sputtering of Zr (circles), and results reported in 2018 by Chakraborty et al. [3] (triangles). The dashed and dash-dotted line in panel a show lattice parameter obtained by averaging of atomic volumes and sizes (Vegard law), respectively. The inset in panel a shows the grain refinement with increasing [Zr] and the inset in panel b shows the correlation of stress-free lattice parameter and hardness, in both cases for DC sputtering of both Cu and Zr.

The measured hardness of Cu-Zr films prepared by the two sputtering techniques is shown in Fig. 1b. The trends are actually very similar to those in Fig 1a: linear increase from ≈2.5 to ≈4 GPa in the [Zr] range from 0 to ≈1%, followed by gradual saturation between ≈1% and ≈3%. In other words, there is a strong correlation of hardness and stress-free lattice parameter and in turn [$Zr_{cr}$], especially in the case of less noisy data obtained by DC sputtering of both Cu and Zr. This correlation (inset in Fig. 1b) can be quantified by a Pearson rank correlation coefficient of 0.97 (within error bars from 1). Collectively, the results strongly indicate that there are two phenomena which could contribute to the hardness enhancement in this [Zr] range: not only the extrinsic hardness given by a grain refinement but also the intrinsic hardness given by solid solution strengthening (see the discussion for other binary Cu-based alloys [2]).

The hardness values obtained by HiPIMS of Cu and DC sputtering of Zr are not only noisier but also of 0.3-1.0 GPa higher. While the lower bound of this difference (0.3 GPa, close to that measured also for Zr-free Cu) can be explained by densification due to the energetic bombardment, the rest of this difference indicates a relatively higher contribution of nanocomposite formation allowed by the higher energy delivered into the growing films and the consequently relatively lower [$Zr_{cr}$]. The presented trends (once again) smoothly connect to those reported in [3], except that the latter (once again) continue to even higher [Zr] values.

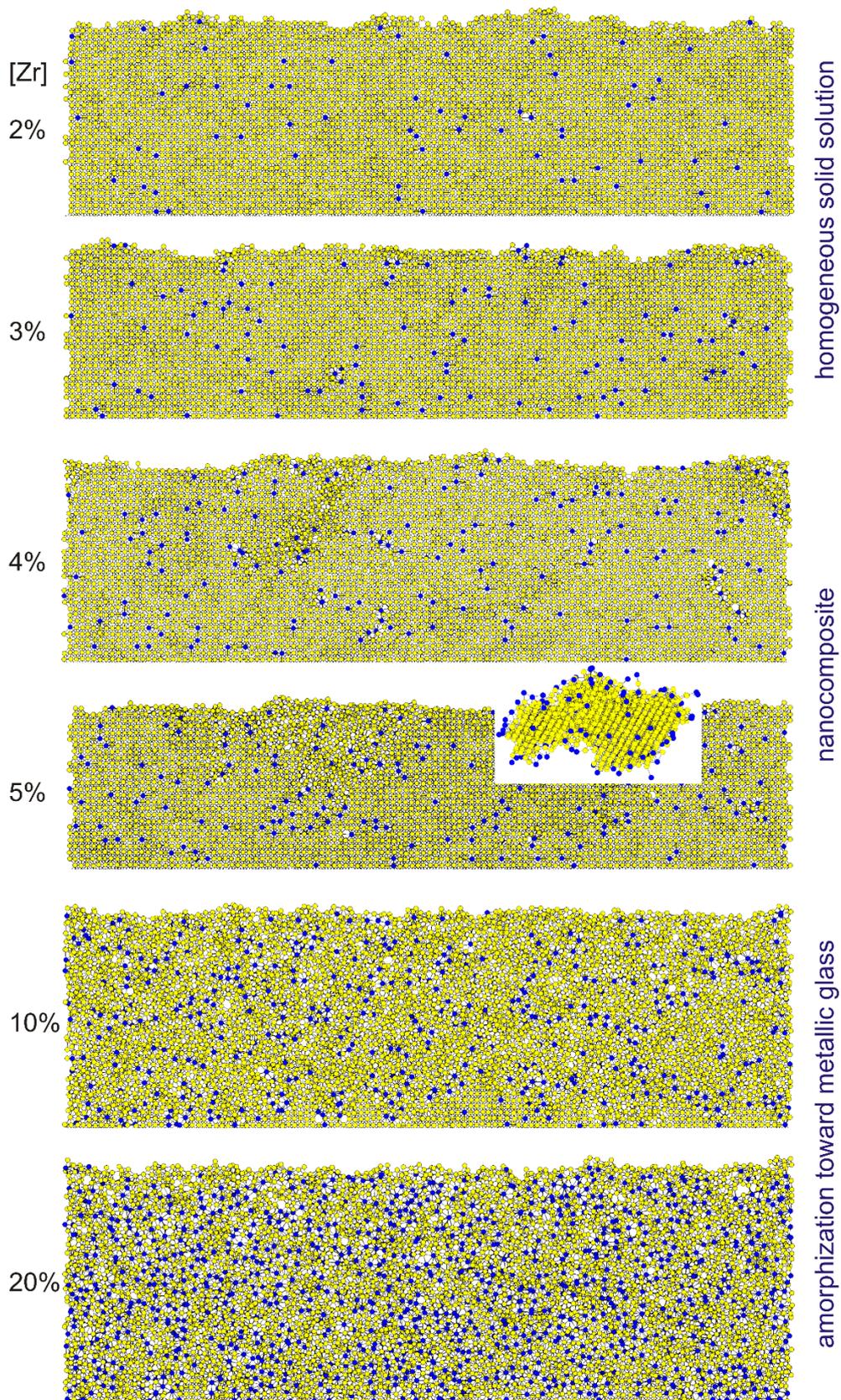

**Figure 2**: Snapshots from the end of MD simulations of the atom-by-atom growth of Cu-Zr. Yellow balls are Cu, blue balls are Zr, and the Cu (001) growth template is not shown. The horizontal dimension is 22 nm. The snapshots obtained at [Zr] = 0 and 1% are similar to that obtained at [Zr] = 2%. One of the Cu crystals surrounded by Zr, obtained at [Zr] = 5%, is enlarged in the inset.

The Cu-Zr structures obtained by growth simulations are shown in Fig. 2, their short- and medium-range order is quantified in Fig. 3 in terms of bonding statistics and statistics of how many bonds far is the nearest Zr atom, and their long-range order is quantified in Fig. 4 in terms of size of the second largest grain (using the definition of boundary atoms given in Sec. 2.2). Note that a large second largest grain is a fingerprint of nanocomposite, while a small second largest grain is a fingerprint of either solid solution (the largest grain constitutes a crystal spanning whole simulation cell) at low [Zr] or amorphization at high [Zr]. A case can be made that the low energy of arriving atoms makes the simulations a counterpart of experiments utilizing DC sputtering of Cu (less noisy squares in Fig. 1) rather than HiPIMS of Cu (more noisy circles in Fig. 1).

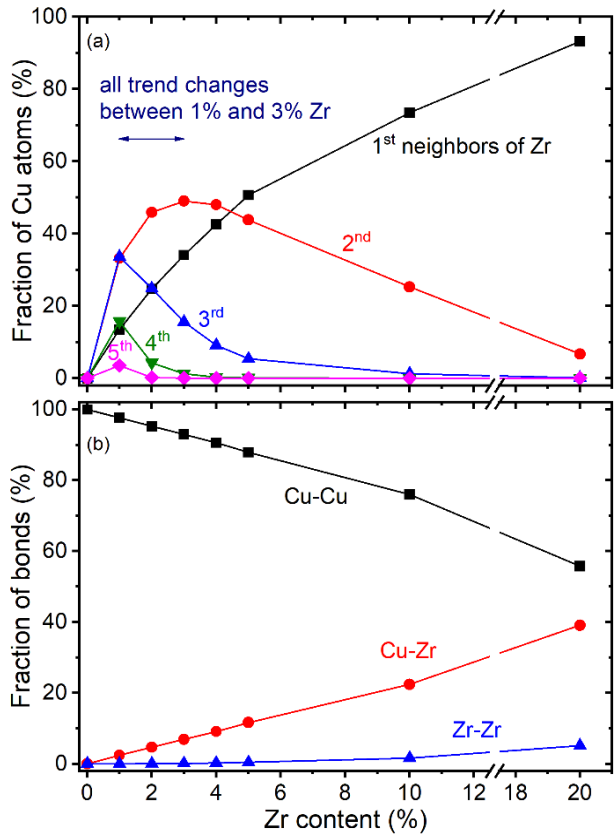

**Figure 3**: Atomic structure of simulated Cu-Zr (Fig. 2) in terms of statistics of Cu atoms which are $n^{th}$ neighbor of Zr (panel a) and bonding statistics (panel b).

The structures obtained at [Zr] ≤ 3% are almost perfect monocrystals which can be described (except pure Cu) as homogeneous supersaturated solid solutions. See the snapshots for [Zr] = 2% and 3% (those for lower [Zr] are very similar) shown in Fig. 2, excluded clustering of Zr due to almost zero number of Zr-Zr bonds shown in Fig. 3b, and very small second largest grain in Fig. 4a. However, there is an important phenomenon which takes place in this compositional interval: almost all changes of statistics of how many bonds far is the nearest Zr atom shown in Fig. 3a. While the number of 1st nearest neighbors of Zr simply increases with [Zr], the numbers of 2nd, 3rd, 4th and 5th nearest neighbors of Zr exhibit maxima (taking the non-symmetrical peak shapes into account) at 3%, between 1 and 2%, slightly above 1% and at 1% of Zr, respectively. Collectively, these findings confirm the interpretation of the corresponding experimental results presented in Fig. 1. On the one hand, there is no evidence for grain boundary segregation which could be correlated with hardness changes. On the other hand, there are changes inside the supersaturated solid solutions which take place in the same compositional interval up to $[Zr] \approx 3\%$ as the hardness changes, thereby explaining most of these hardness changes. This is not to question the aforementioned grain refinement taking place at a scale larger than the simulation cell size and caused also or mainly by the kinetics of the metastable supersaturated crystal growth rather than only by a grain boundary segregation.

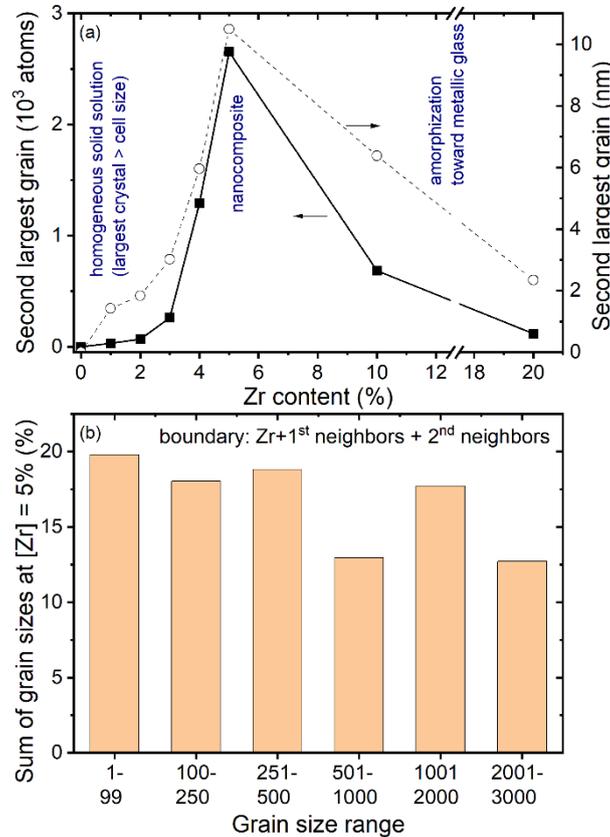

**Figure 4**: Crystallinity of simulated Cu-Zr (Fig. 2), assuming that boundaries between crystals are formed by Zr atoms, their 1st neighbors and 2nd neighbors. Panel a shows the size of second largest crystal: small in the case of formation of homogeneous solid solution spanning whole simulation cell or in the case of amorphization, but large in the case of nanocomposite formation. Panel b shows the statistics of grain sizes at $[Zr] = 5\%$, obtained using the aforementioned strict definition (leading to relatively small grains) and for completeness presented down to the zones smaller than hundred atoms (not necessarily grains in a narrower sense of the word).

The structures obtained at $[Zr] = 4\text{-}5\%$ are nanocomposites, largely consisting of Cu-based nanocrystals. See the cross-section in the corresponding panels of Fig. 2, and the extremal size of the second largest nanocrystal in Fig. 4a. The amount of Zr segregated at nanocrystal surfaces is already high enough to completely envelope the nanocrystals by boundary atoms as defined in Sec. 2.2: Zr + 1st neighbor of Zr + 2nd neighbors of Zr (but not by only Zr - Fig. 2 clearly shows that such an imagination would be too idealized). Note that not all boundary atoms in the present sense of the word are at the very nanocrystal surface: the surface layer defined in this way is up to three monolayers thick. Only in this context it is not surprising that, as shown in Fig. 3a, most atoms in these compositions are boundary atoms. The most typical nanocomposite structure formed at $[Zr] = 5\%$ can be quantified by sizes of the largest (not shown) and second largest (shown) nanocrystal slightly below 3 000 atoms or $\approx 10$ nm in the longest direction (Fig. 4a), and by comparable numbers of atoms in grains belonging to size groups 2001-3000 atoms, 1001-2000 atoms, 501-1000 atoms, etc. (Fig. 4b). One of the nanocrystals is enlarged in Fig. 2 as an inset, and a movie showing this crystal from all directions is provided as a Supplementary material.

It is also worth to point out the parallel of the presented structural evolution to that known for thoroughly studied nitride nanocomposite-forming systems such as TiN/SiN$_x$ [39]: possibility to form supersaturated TiN-based solid solutions containing 8-9% Si in the cation sublattice (4-4.5% overall) in both experiments [40] and liquid-quench simulations [41], and formation of nanocomposites containing nanometer-size crystals at 10-20% Si in the cation sublattice (5-10% overall). The same effects are achieved by lower [Zr] in Cu-Zr than [Si] in Ti-Si-N, arguably because Zr is significantly larger than Cu while the sizes of Ti and Si are comparable.

At [Zr] increasing to ≥ 10%, the materials gradually convert from a nanocomposite to a metallic glass. Although the materials are still not fully homogeneous on the atomic scale, the size of Cu-rich grains drops below 3 nm ([Zr] = 10%) and next below 1 nm ([Zr] = 20%), leading to the grain amorphization. Here, we are leaving the scope of the present paper: the Cu-Zr metallic glasses prepared by us are studied by experiments and growth simulations in detail elsewhere [24].

**Conclusions**

The non-equilibrium growth of Cu-rich Cu-Zr thin films has been investigated by a combination of magnetron sputter deposition and molecular dynamics simulations. The incorporation of 3% Zr allowed us to increase the film hardness from 2.5 to 4 GPa (DC sputtering of both Cu and Zr) or from 3 to 5 GPa (HiPIMS of Cu + DC sputtering of Zr). The stress-free lattice parameters and the growth simulations collectively reveal that a significant part of Zr atoms are in this range in the supersaturated solid solution rather than at the grain boundaries, and that the grain refinement thus has to be explained by lower growth rate of Zr-containing crystals as such rather than by the grain boundary segregation. The incorporation of Zr into supersaturated solid solution correlates with the hardness enhancement equally well as the grain size. Further growth simulations reveal a transition from a solid solution (0-3% Zr) through a nanocomposite (4-5% Zr) to a metallic glass (10-20% Zr), which is not accompanied by a further hardness enhancement. From the methodology point of view, the simulations constitute an early example of modelling the atom-by-atom nanocomposite growth, and the processing of their results illustrates the necessity of realistic representation of their boundaries (Zr atoms and their 1$^{st}$ and 2$^{nd}$ nearest neighbors rather than only Zr).

**Acknowledgment**

This work was supported by the Czech Science Foundation under the project No. 23-07924I and by the project QM4ST, funded as project No. CZ.02.01.01/00/22_008/0004572 by P JAC, call Excellent Research. Computational resources were provided by the e-INFRA CZ project (ID:90254), supported by the MEYS of the Czech Republic.

**Data availability**

Data will be made available on request.

**References**

1. P. Zhang, J.Y. Zhang, J. Li, G. Liu, K. Wu, Y.Q. Wang, J. Sun, Microstructural evolution, mechanical properties and deformation mechanisms of nanocrystalline Cu thin films alloyed with Zr, Acta Mater. 76 (2014) 221-237, doi.org/10.1016/j.actamat.2014.04.041.
2. J.T. Zhao, J.Y. Zhang, L.F. Cao, Y.Q. Wang, P. Zhang, K. Wu, G. Liu, J. Sun, Zr alloying effect on the microstructure evolution and plastic deformation of nanostructured Cu thin films, Acta Mater. 132 (2017) 550-564, doi.org/10.1016/j.actamat.2017.05.007.
3. J. Chakraborty, T. Oellers, R. Raghavan, A. Ludwig, G. Dehm, Microstructure and residual stress evolution in nanocrystalline Cu-Zr thin films, J. Alloys Compd 896 (2022) 162799, doi.org/10.1016/j.jallcom.2021.162799.
4. P. Zeman, M. Zitek, S. Zuzjakova, R. Cerstvy, Amorphous Zr-Cu thin-film alloys with metallic glass behavior, J. Alloy. Compd. 696 (2017) 1298-1306, doi.org/10.1016/j.jallcom.2016.12.098.
5. J. Dudonis, R. Brucas, A. Miniotas, Synthesis of amorphous Zr-Cu alloys by magnetron co-sputtering, Thin Solid Films 275 (1996) 164-167, doi.org/10.1016/0040-6090(95)07033-8.
6. M. Apreutesei, P. Djemia, L. Belliard, G. Abadias, C. Esnouf, A. Billard, P. Steyer, Structural-elastic relationships of Zr-TL (TL = Cu, Co, Ni) thin films metallic glasses. J. Alloy. Compd. 707 (2017) 126-131, doi.org/10.1016/j.jallcom.2016.12.208.
7. M.A. Atwater, R.O. Scattergood, C.C. Koch, The stabilization of nanocrystalline copper by zirconium, Mater. Sci. Eng. A 559 (2013) 250-256, doi.org/10.1016/j.msea.2012.08.092.
8. A. Khalajhedayati, T.J. Rupert, High-temperature stability and grain boundary complexion formation in a nanocrystalline Cu-Zr alloy, JOM 67 (2015) 2788-2801, doi.org/10.1007/s11837-015-1644-9.
9. A.I. Zubkov, E.N. Zubarev, O.V. Sobol, M.A. Hlushchenko, E.V. Lutsenko, Structure of vacuum Cu-Ta condensates, Phys. Met. Metallogr. 118 (2017), 158-163, doi.org/10.1134/S0031918X17020156.
10. M.A. Zhadko, A.I. Zubkov, Effect of molybdenum on the structure and strength of copper vacuum condensates, In: V. Tonkonogyi et al., Advanced Manufacturing Processes. InterPartner 2019. Lecture Notes in Mechanical Engineering. Springer, Cham, doi.org/10.1007/978-3-030-40724-7_49.
11. J.E. Greene, Review Article: Tracing the recorded history of thin-film sputter deposition: From the 1800s to 2017, J. Vac. Sci. Technol. A 35 (2017) 05C204, doi.org/10.1116/1.4998940.
12. F.V. Grigoriev, V.B. Sulimov, Atomistic simulation of physical vapor deposition of optical thin films, Nanomaterials 13 (2017) 1717, doi.org/10.3390/nano13111717.
13. J. Houska, S. Mraz, and J. M. Schneider, Experimental and molecular dynamics study of the growth of crystalline $TiO_2$, J. Appl. Phys. 112 (2012) 073527, doi.org/10.1063/1.4757010.
14. J. Houska, Molecular dynamics study of the growth of crystalline $ZrO_2$, Surf. Coat. Technol. 304 (2016) 23-30, doi.org/10.1016/j.surfcoat.2016.07.004.
15. J. Houska, Pathway for a low-temperature deposition of α-$Al_2O_3$: a molecular-dynamics study, Surf. Coat. Technol. 235 (2013) 333-341, doi.org/10.1016/j.surfcoat.2013.07.062.
16. J. Houska, P. Zeman, Role of Al in Cu-Zr-Al thin film metallic glasses: molecular dynamics and experimental study, Comp. Mater. Sci. 222 (2023) 112104, doi.org/10.1016/j.commatsci.2023.112104.

17. N.C. Cooper, M.S. Fagan, C.M. Goringe, N.A. Marks, D.R. McKenzie, Surface structure and sputtering in amorphous carbon thin films: a tight-binding study of film deposition, J. Phys.: Condens. Matter 14 (2002) 723-730, doi.org/10.1088/0953-8984/14/4/307.
18. H. Hensel, H.M. Urbassek, Simulation of the influence of energetic atoms on Si homoepitaxial growth, Phys. Rev. B 58 (1998) 2050-2054, doi.org/10.1103/PhysRevB.58.2050.
19. V.I. Ivashchenko, P.E.A. Turchi, V.I. Shevchenko, O.A. Shramko, Molecular dynamics simulations of a-SiC films, Phys. Rev. B 70 (2004) 115201, doi.org/10.1103/PhysRevB.70.115201.
20. J. Houska, J.E. Klemberg-Sapieha, L. Martinu, Atom-by-atom simulations of chemical vapor deposition of nanoporous hydrogenated silicon nitride, J. Appl. Phys. 107 (2010) 083501, doi.org/10.1063/1.3371680.
21. M. Saraiva, V. Georgieva, S. Mahieu, K. Van Aeken, A. Bogaerts, D. Depla, Compositional effects on the growth of Mg(M)O films, J. Appl. Phys. 107 (2010) 034902, doi.org/10.1063/1.3284949.
22. K. Hantova, J. Houska, Molecular dynamics study of the growth of $ZnO_x$ films, J. Appl. Phys. 132 (2022) 185304, doi.org/10.1063/5.0106856.
23. L. Xie, H. An, Q. Peng, Q. Qin, Y. Zhang, Sensitive Five-Fold Local Symmetry to Kinetic Energy of Depositing Atoms in Cu-Zr Thin Film Growth, Materials 11 (2018) 2548, doi.org/10.3390/ma11122548.
24. J. Houska, P. Machanova, M. Zitek, P. Zeman, Molecular dynamics and experimental study of the growth, structure and properties of Zr-Cu films, J. Alloys Compd. 828 (2020) 154433, doi.org/10.1016/j.jallcom.2020.154433.
25. G.B. Bokas, L. Zhao, D. Morgan, I. Szlufarska, Increased stability of CuZrAl metallic glasses prepared by physical vapor deposition, J. Alloys Comp. 728 (2017) 1110-1115, doi.org/10.1016/j.jallcom.2017.09.068.
26. Y.J. Hu, Y. Wang, W.Y. Wang, K.A. Darling, L.J. Kecskes, Z.K. Liu, Comp. Mater. Sci. 171 (2020) 109271, doi.org/10.1016/j.commatsci.2019.109271.
27. M. Wagih, C.A. Schuh, Acta Mater. 217 (2021) 117177, doi.org/10.1016/j.actamat.2021.117177.
28. G.J. Yang, B. Xu, L.T. Kong, J.F. Li, S. Zhao, Size effects in $Cu_{50}Zr_{50}$ metallic glass films revealed by molecular dynamics simulations, J. Alloys Compd. 688 (2016) 88-95, doi.org/10.1016/j.jallcom.2016.07.178.
29. Z.D. Sha, Y.W. Zhang, Y.P. Feng, Y. Li, Molecular dynamics studies of short to medium range order in $Cu_{64}Zr_{36}$ metallic glass, J. Alloys Compd. 509 (2011) 8319-8322, doi.org/10.1016/j.jallcom.2011.05.071.
30. Y.Q. Cheng, H.W. Sheng, E. Ma, Relationship between structure, dynamics, and mechanical properties in metallic glass-forming alloys, Phys. Rev. B 78 (2008) 014207, doi.org/10.1103/PhysRevB.78.014207.
31. Z.D. Sha, Y.P. Feng, Y. Li, Statistical composition-structure-property correlation and glass-forming ability based on the full icosahedra in Cu-Zr metallic glasses, Appl. Phys. Lett. 96 (2010) 061903, doi.org/10.1063/1.3310278.
32. N. Mattern, P. Jóvári, I. Kaban, S. Gruner, A. Elsner, V. Kokotin, H. Franz, B. Beuneu, J. Eckert, Short-range order of Cu–Zr metallic glasses, J. Alloys Compd. 485 (2009) 163-169, doi.org/10.1016/j.jallcom.2009.05.111.
33. C.D. Wu, H.X. Li, Molecular dynamics simulation of strengthening of nanocrystalline Cu alloyed with Zr, Mater. Today Commun. 26 (2021) 101963, doi.org/10.1016/j.mtcomm.2020.101963.
34. S.J. Plimpton, Fast Parallel Algorithms for Short-Range Molecular Dynamics, J. Comp. Phys. 117 (1995) 1-19, doi.org/10.1006/jcph.1995.1039, lammps.sandia.gov.

35 M.I. Mendelev, D.J. Sordelet, M.J. Kramer, J. Appl. Phys. 102 (2007) 043501, doi.org/10.1063/1.2769157.
36 Y. Hu, J.D. Schuler, T.J. Rupert, Identifying interatomic potentials for the accurate modeling of interfacial segregation and structural transitions, Comp. Mater. Sci. 148 (2018) 10-20, doi.org/10.1016/j.commatsci.2018.02.020.
37 J.F. Panzarino, T.J. Rupert, Tracking Microstructure of Crystalline Materials: A Post-Processing Algorithm for Atomistic Simulations, JOM 66 (2014) 417-428, doi.org/10.1007/s11837-013-0831-9.
38 M.J. Tenwick, H.A. Davies, Enhanced Strength in High Conductivity Copper Alloys, Mater. Sci. Eng. 98 (1988) 543-546, doi.org/10.1016/0025-5416(88)90226-1.
39 S. Veprek, S. Reiprich, A concept for the design of novel superhard coatings, Thin Solid Films 268 (1995) 64-71, doi.org/10.1016/0040-6090(95)06695-0.
40 A. Flink, M. Beckers, J. Sjölén, T. Larsson, S. Braun, L. Karlsson. L. Hultman, The location and effects of Si in $(Ti_{1-x}Si_x)N_y$ thin films, J. Mater. Res. 24 (2009) 2483-2498, doi.org/10.1557/jmr.2009.0299.
41 J. Houska, J.E. Klemberg-Sapieha, L. Martinu, Atomistic simulations of the characteristics of TiSiN nanocomposites of various compositions, Surf. Coat. Technol. 203 (2009) 3348-3355, doi.org/10.1016/j.surfcoat.2009.04.021.